\documentclass{iopart}

\usepackage{epsfig}

\begin{document}

\title{Hadronization in heavy ion collisions: recombination or fragmentation?}

\author{R~J~Fries\dag, B~M\"uller\dag, C~Nonaka\dag\ and S~A~Bass\dag\ddag}

\address{\dag\ Physics Department, Duke University, P.O.Box 90305, Durham, 
  NC 27708, USA}
\address{\ddag\ RIKEN BNL Research Center, Brookhaven National Laboratory,
  Upton, NY 11973, USA}

\ead{rjfries@phy.duke.edu}

\begin{abstract}
  We show that hadron production in relativistic heavy ion collisions
  at transverse momenta larger than 2 GeV/$c$ can be explained by the
  competition of two different hadronization mechanisms. Above 
  5 GeV/$c$ hadron production can be described by
  fragmentation of partons that are created perturbatively. Below 5 GeV/$c$
  recombination of partons from the dense and hot fireball dominates.
  This can explain some of the surprising features of RHIC data like
  the constant baryon-to-meson ratio of about one and the small nuclear
  suppression for baryons between 2 to 4 GeV/$c$.
\end{abstract}

\submitto{JPG}


The Relativistic Heavy Ion collider (RHIC) has provided exciting data
about hadron production at transverse momenta of a few GeV/$c$ in central
Au+Au collisions. The production of pions at high $P_T$ was found to be 
suppressed compared to the scaled yield from $p+p$ collisions \cite{PHENIX}. 
This jet quenching effect can be understood by final state interaction 
of fast partons with the dense and hot 
medium produced in central heavy ion collisions. Fast partons lose energy 
via induced bremsstrahlung before they can fragment into high $P_T$ hadrons 
\cite{GyulWang:94}. The suppression effect is dramatic and can be as large
as a factor of 6 above $P_T=5$ GeV/$c$.

On the other hand the suppression of protons and antiprotons seems to be 
much less \cite{PHENIX-B}. Experimental data from PHENIX show
a proton/pion ratio of about 1 between 1.5 GeV/$c$ and 4 GeV/$c$
\cite{Chujo:02}. 
This is surprising since the production of protons and antiprotons is usually
suppressed compared to the production of pions because of the much larger mass.
At high transverse momentum this can be understood in terms
of perturbative quantum chromodynamics (pQCD) \cite{CoSo:81}. 
The fragmentation functions $D_{a\to h}(z)$ describe the probability for a 
parton $a$ with momentum $p$ to turn into a hadron with momentum 
$zp$, $0<z<1$. These fragmentation functions were measured for pions and
protons, mainly in $e^+ e^-$ collisions, and give a ratio 
$p/\pi^0 < 0.2$ for $P_T > 2$ GeV/$c$ when used in $p+p$ and $N+N$ 
collisions. The energy loss of partons in a medium can be taken into account 
by a rescaling of the parton momentum \cite{GW:00}. However this should
affect baryons and mesons in the same way.

The lack of nuclear suppression for baryons is a challenge for our 
understanding of hadron production. The currently accepted picture assumes
that a parton with large transverse momentum is produced in a hard scattering 
reaction between initial partons, propagates through the surrounding hot 
matter and loses energy by interactions, and finally hadronizes. Apparently 
the creation and interaction of a parton will happen independently of
its later fate during hadronization. Therefore any unusual behavior that
is different between baryons and mesons can only be attributed to
hadronization itself. We propose to use recombination of quarks from
the surface of the hot fireball as an alternative hadronization
mechanism \cite{FMNB:03}.

In the fragmentation process a parton with transverse momentum $p_T$ is 
leaving the interaction zone while still being connected with other partons 
by a color string. The breaking of the
string creates quark antiquark pairs which finally turn into hadrons.
The distribution of one of these hadrons, which is bound to have less 
transverse momentum $P_T = zp_T$, is then described by a fragmentation 
function. The average value $\langle z\rangle$ is about 0.5 for pions
in $p+p$ collisions. In other words the production of a, say, 5 GeV/$c$
pion has to start with a 10 GeV/$c$ parton in average, which are rare to find
due to 
the steeply falling parton spectrum. Jet quenching even enhances the lack of 
high $p_T$ partons. On the other hand, the 5 GeV/$c$ pion could be produced by
the recombination of a quark and an antiquark with about 2.5 GeV/$c$ each in 
average. 2.5 GeV/$c$ and 10 GeV/$c$ are separated by orders of magnitude in 
the parton spectrum. The price to pay is of course that two of these partons
have to be found close to each other in phase space. However we do have a 
densely populated phase space in central heavy ion collisions at RHIC where 
we even expect the existence of a thermalized quark gluon plasma. 

Recombination of quarks has been considered before in hadron
collisions \cite{DasHwa:77} and
was also applied to heavy ion collisions \cite{Gupt:83}. In QCD the 
leading particle effect in the forward region of a hadron collision is well
known. This is the phenomenon that the production of hadrons that share
valence partons with the beam hadrons are favored in forward direction. It 
has been realized that this can only be explained by recombination. This has 
fueled a series of theoretical work, see e.g. \cite{BJM:02} and references 
therein. Recently the recombination idea for heavy ion collisions, stimulated 
by the RHIC results, has been revived for elliptic flow 
\cite{Voloshin:02} and hadron spectra and ratios 
\cite{FMNB:03,GreKoLe:03}.

The formalism of recombination has already been developed in a 
covariant setup for the process of baryons coalescing into light nuclei and 
clusters in nuclear collisions \cite{DHSZ:91,ScheiHei:99}.
We give a brief derivation for the case of mesons.
By introducing the density matrix $\hat \rho$ for the system of partons, 
the number of quark-antiquark states that will be measured as mesons is given 
by
\begin{equation}
  \label{eq:pinumb}
  N_M = \sum_{ab}\int \frac{d^3 P}{(2\pi)^3} \> \langle M ;{\bf P} | \> \hat 
  \rho_{ab} \> | M ;{\bf P} \rangle
\end{equation}
Here $| M ;{\bf P} \rangle$ is a meson state with momentum ${\bf P}$ and the
sum is over all combinations of quantum numbers --- flavor, helicity and 
color --- of valence partons that contribute to the given meson $M$.
This can be cast in covariant form using a hypersurface $\Sigma$ for
hadronization \cite{CoFr:74}
\begin{equation}  \fl
  E \frac{d N_M}{d^3 P} = {C_M} \int\limits_\Sigma 
  \frac{d \sigma \> P\cdot u(\sigma)}{(2\pi)^3}  \int \frac{d^3 q
  }{(2\pi)^3} \>  
  w_a\bigg( {\sigma} ; \frac{\bf P}{2}-{\bf q} \bigg)
   \> \Phi^W_M ({\bf q}) \>
  w_b\bigg( {\sigma}; \frac{\bf P}{2}+{\bf q}
  \bigg).
\end{equation}
Here $E$ is the energy of the four vector $P$, $d \sigma$ a measure on $\Sigma$
and $u(\sigma)$ is the future oriented unit vector orthogonal to the 
hypersurface $\Sigma$. $w_a$ and $w_b$ are the phase space densities for 
the two partons $a$ and $b$, $C_M$ is a degeneracy factor and 
\begin{equation} 
  \Phi^W_M ({\bf q}) = \int d^3 r \> \Phi^W_M ({\bf r},{\bf q})
\end{equation}
where $\Phi^W_M ({\bf r},{\bf q})$ is the Wigner function of the meson 
\cite{ScheiHei:99}.

Since the hadron structure is best known in a light cone frame, we write
the integral over ${\bf q}$ in terms of light cone coordinates in a frame
where the hadron has no transverse momentum but a large light cone component
$P^+$. This can be achieved by a simple rotation from the lab frame.
Introducing the momentum $k=P/2-q$ of parton $a$ in this frame we have
$k^+ = x P^+$ with $0<x<1$. We make an ansatz for the Wigner function of the
meson in terms of light cone wave functions $\phi_M(x)$. The final result
can be written as \cite{FMNB:03}
\begin{equation} \fl
  \label{eq:res3}
  E \frac{N_M}{d^3 P} = C_M \int\limits_\Sigma d\sigma 
  \frac{P\cdot u(\sigma)}{(2\pi)^3} \int\limits_0^1  {d x} \>  
  w_a\big( {\sigma} ; x {P^+} \big)
   \> \left| \phi_M (x) \right|^2 \>
  w_b\big( {\sigma}; (1-x) {P^+} \big).
\end{equation}
For a baryon with valence partons $a$, $b$ and $c$ we obtain
\begin{eqnarray}
  \label{eq:protres}
  E \frac{N_B}{d^3 P} &= C_B \int\limits_\Sigma d\sigma 
  \frac{P\cdot u(\sigma)}{(2\pi)^3} \int\limits_0^1  {d x_1 \, d x_2 \, d x_3}
  \delta(x_1+x_2+x_3-1) \\  & \times
  w_a\big( {\sigma} ; x_1 {P^+} \big)
  w_b\big( {\sigma}; x_2 {P^+} \big) 
  w_c\big( {\sigma}; x_3 {P^+} \big)
   \> \left| \phi_B (x_1,x_2,x_3) \right|^2 . \nonumber
\end{eqnarray}
$\phi_B(x_1,x_2,x_3)$ is the effective wave function of the baryon in light 
cone coordinates. 

A priori these wave functions are not equal to the 
light cone wave functions used in exclusive processes. We are recombining
effective quarks in a thermal medium and not perturbative partons in an 
exclusive process.
Nevertheless, as an ansatz for a realistic wave function one can adopt
the asymptotic form of the light cone distribution amplitudes
\begin{eqnarray}
  \label{eq:barlcwf}
  \phi_M(x) &= \sqrt{30} x(1-x), \\
  \phi_B(x_1,x_2,x_3) &= 12\sqrt{35} \, x_1 x_2 x_3 \nonumber
\end{eqnarray}
as a model.
However, it turns out that for a purely exponential spectrum the shape of the 
wave function does not matter. In that case the dependence on $x$ drops out
of the product of phase space densities
\begin{equation}
  w_a\big( {\sigma} ; x {P^+} \big) w_b\big( {\sigma}; (1-x) {P^+} \big)
  \sim e^{-xP^+/T} e^{-(1-x)P^+/T} = e^{-P^+/T}.
\end{equation}
One can show that a narrow width approximation, using $\delta$ peaked
wave functions that distribute the momentum of the hadron equally among
the valence quarks --- 1/2 in the case of a meson and 1/3 in the
case of a baryon --- deviates by less then 20\% from a calculation using
the wave functions in (\ref{eq:barlcwf}).
That there is a small deviation can be attributed to the violation of energy
conservation. Since recombination is a $2\to 1$ or $3\to 1$ process,
energy will generally not be conserved if we enforce momentum conservation and 
a mass shell condition for all particles. However, one can show that violations
of energy conservation are of the order of the effective quark
masses or $\Lambda_{\rm QCD}$ and can therefore be neglected for transverse
hadron momenta larger than 2 GeV/$c$.

Fragmentation of partons is given by \cite{Owens:86}
\begin{equation}
  \label{eq:fracmaster}
  E \frac{d \sigma_h}{d^3 P} = \sum_a \int\limits_0^1 \frac{d z}{z^2} 
  D_{a\to h}(z) E_a \frac{d \sigma_a}{d^3 P_a}.
\end{equation}
The sum runs over all parton species $a$ and $\sigma_a$ is the cross section
for the production of parton $a$ with momentum $P_a = P/z$.
We use a leading order (LO) pQCD calculation of $\sigma_a$ \cite{FMS:02}
together with LO KKP fragmentation functions \cite{KKP:00}. Energy loss
of the partons is taken into account by a shift of the parton spectrum by
\begin{equation}
  \Delta p_T = \sqrt{ \lambda p_T}.
\end{equation}

We summarize that the transverse momentum dependent yield of mesons from 
recombination can be written as $\sim C_M w^2(P_T/2)$ in the simple narrow 
width approximation, whereas from fragmentation we expect 
$\sim D(z) \otimes w(P_T/z)$. For an exponential parton spectrum 
$w=e^{-P_T/T}$ the ratio of recombination to fragmentation is
\begin{equation}
  \frac{R}{F} = \frac{C_M}{\langle D \rangle} 
  e^{-\frac{P_T}{T} \left( 1- \frac{1}{\langle z \rangle} \right) }
\end{equation}
where $\langle D \rangle <1$ and $\langle z \rangle <1 $ are average values
of the fragmentation function and the scaling variable. Therefore
$R/F > 1$. In fact, recombination always wins over fragmentation from an 
exponential spectrum (as long as the exponential is not suppressed by 
small fugacity factors). The same is true in the case of baryons.

Now let us consider a power law spectrum $w=A (P_T/\mu)^{-\alpha}$ with
a scale $\mu$ and $\alpha >0 $. Then the ratio of recombination over 
fragmentation is
\begin{equation}
  \frac{R}{F} = \frac{C_M A}{\langle D \rangle} 
  \left(\frac{4}{\langle z \rangle} \right)^\alpha 
  \left(\frac{P_T}{\mu}\right)^{-\alpha}
\end{equation}
and fragmentation ultimately has to win at high $P_T$. 
We also note that we can expect a constant baryon/meson ratio from
recombination, when the parton spectrum is exponential. 
In this case the ratio is only determined by the degeneracy factors
\begin{equation}
  \frac{dN_B}{dN_M} = \frac{C_B}{C_M}.
\end{equation}

For our numerical studies we consider fragmentation of perturbative
partons and recombination from a thermal phase
\begin{equation}
  w_{\rm th} = e^{-p_T \cosh(\eta-y)/T \, e^{-y^2/2 \Delta^2}}
\end{equation}
with an effective temperature $T$. $\eta$ is the space-time rapidity and
$\Delta \approx 2$ the width of the rapidity distribution. We fix the 
hadronization hypersurface $\Sigma$ by the condition 
$\tau_f = \sqrt{t^2-z^2} ={\rm const}.$ \cite{DHSZ:91}. We set 
$\tau_f=5$ fm/$c$. A two phase parton spectrum with a perturbative power
law tail and an exponential part at low transverse momentum is also
predicted by parton cascades like {\sc VNI/BMS} \cite{PCM}.

The parameters, $\lambda$ for the average energy loss and $T$ for the slope
of the exponential parton spectrum, are determined by a fit to the
inclusive charged hadron spectrum measured by PHENIX \cite{Jia:02}.
We obtain $\lambda\approx 1$ GeV and $T\approx 350$ MeV. The temperature
contains the effect of a blue-shift because of the strong radial flow.
An additional normalization factor of about $1/30$ for the recombination part 
is necessary to describe the data. This is due to the use of an effective 
temperature which gives a too large particle number compared to the physical
temperature that we expect to be around 175 MeV.

\begin{figure}
\begin{center}
  \epsfig{file=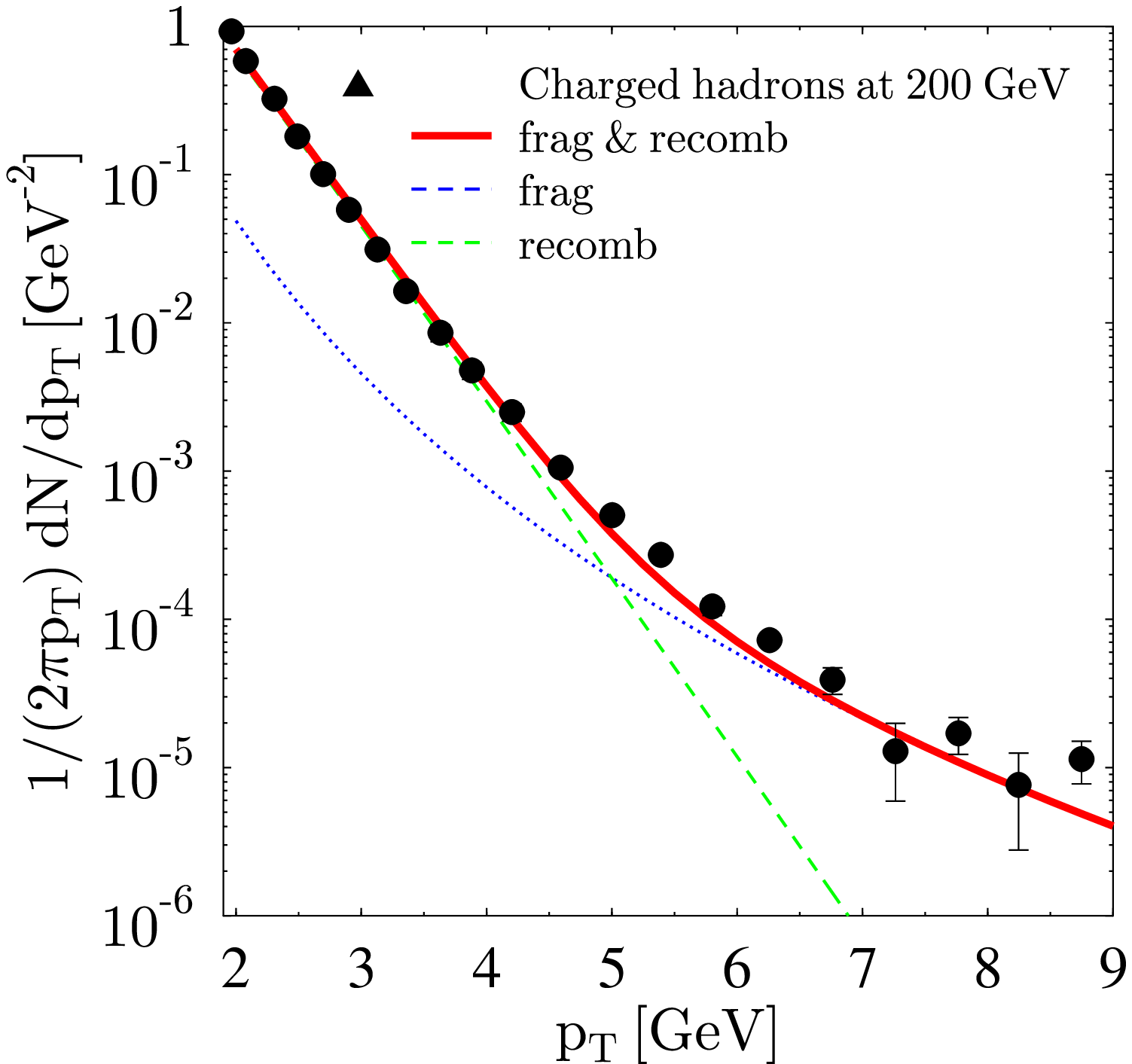,width=6.4cm}
  \epsfig{file=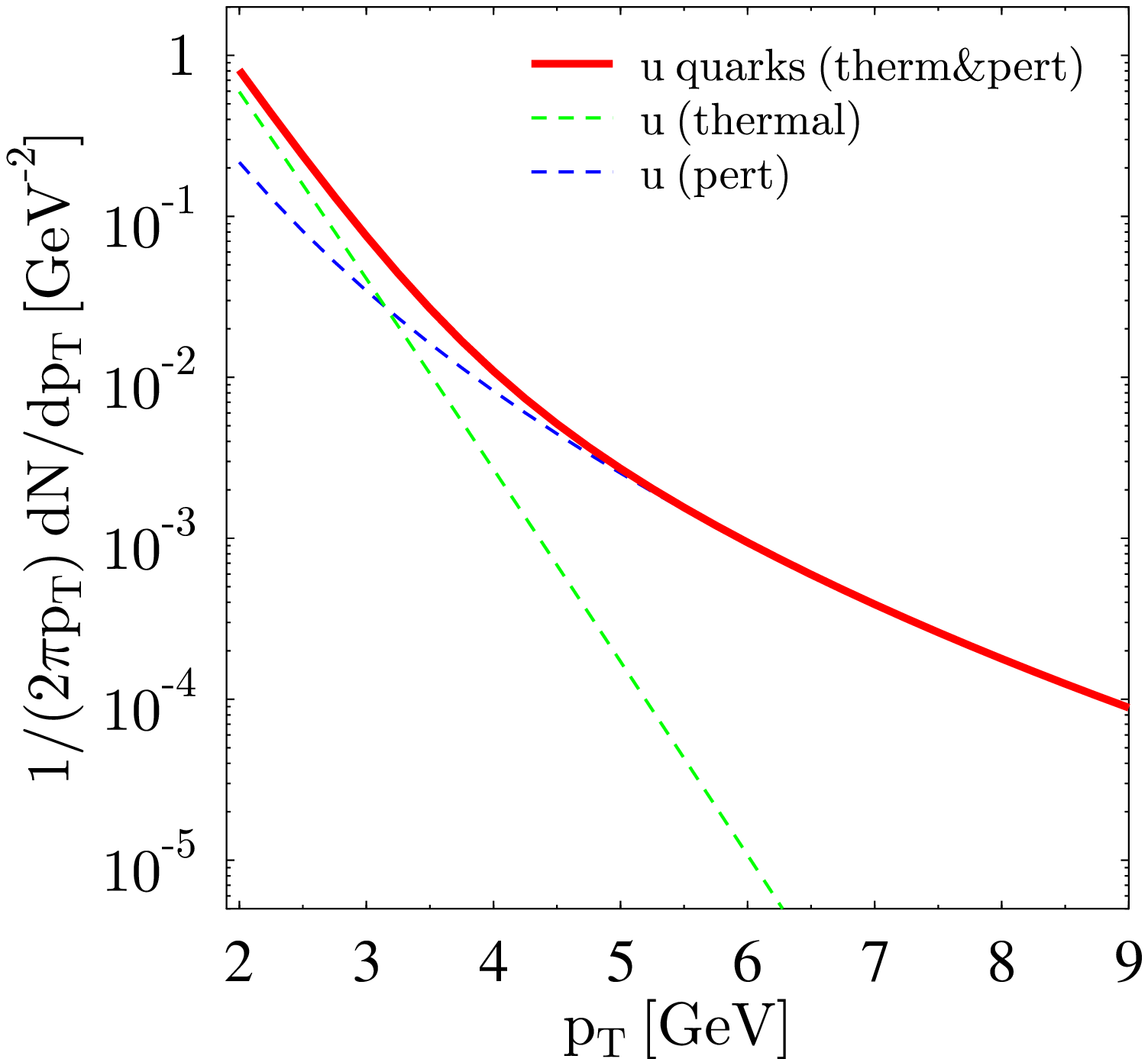,width=6.4cm}  
  \caption{\label{fig:2} Left: the charged hadron spectrum in 
   Au+Au collisions at $\sqrt{S}=200$ GeV as a function of $P_T$. 
   Contributions from fragmentation (dashed), recombination 
   (dotted) and the sum of both (solid line) are shown. Data are from the 
   PHENIX collaboration \cite{Jia:02}. Right: the spectrum of up quarks.}
\end{center}
\end{figure}

In Figure \ref{fig:2} we show the inclusive charged hadron spectrum
using recombination and fragmentation for pions, kaons, protons and 
antiprotons.
We note that the crossover between the recombination dominated and 
fragmentation dominated regimes is around 5 GeV/$c$. It is in fact earlier
for pions ($\sim$4 GeV/$c$) than for protons ($\sim$6 GeV/$c$). 
In Figure \ref{fig:2} we also give the spectrum of up quarks as an example for
the partonic input of our calculation. We note that the crossover between the
perturbative and the thermal domain is here around 3 to 3.5 GeV/$c$. 
It is characteristic for recombination that features of the parton
spectrum are pushed to higher transverse momentum in the hadron spectrum.

\begin{figure}
\begin{center}
  \epsfig{file=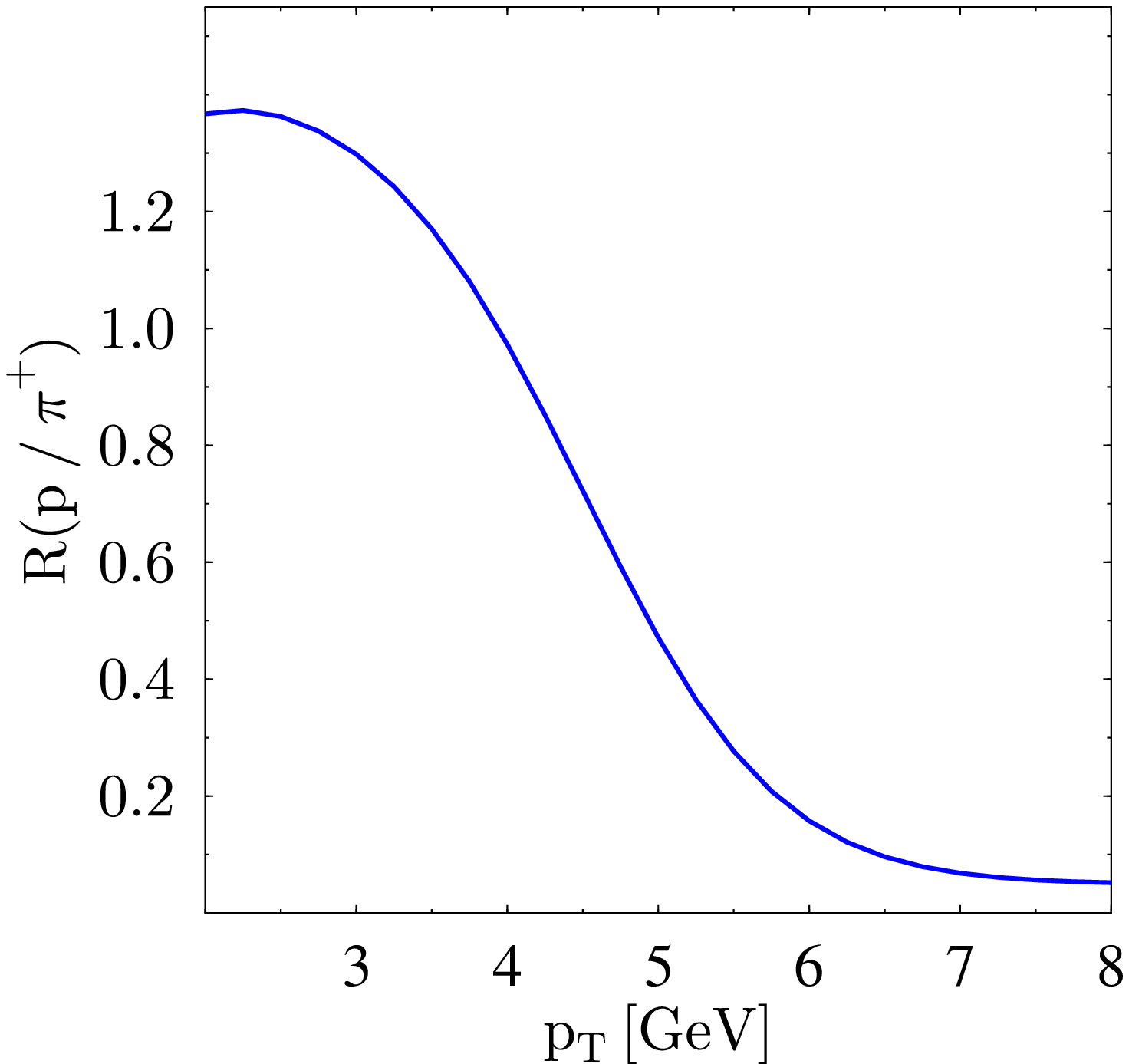,width=6.4cm}
  \epsfig{file=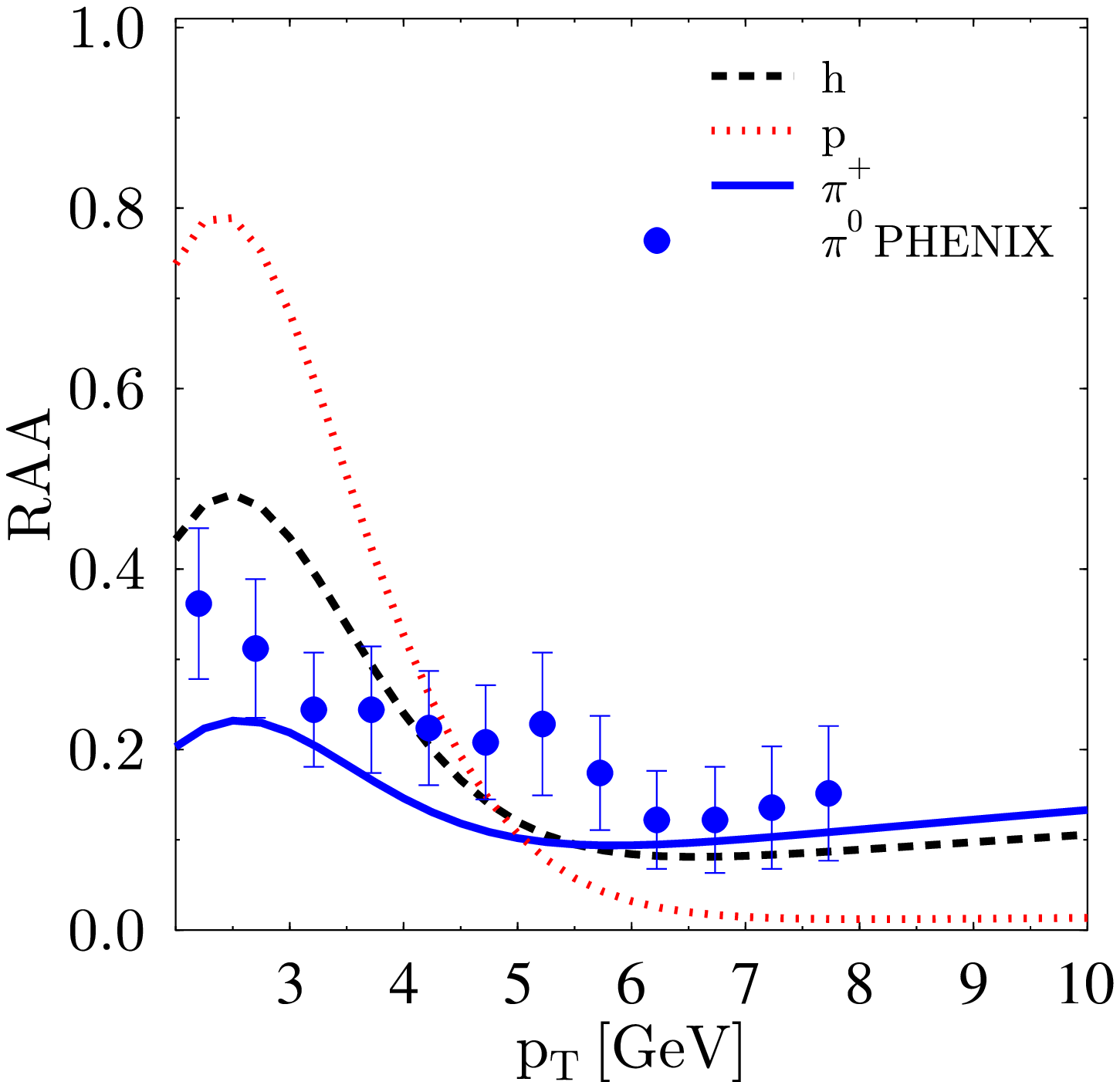,width=6.4cm}
  \caption{\label{fig:3} Left: the
   $p/\pi^+$ ratio exhibits a transition between 4 and 6 GeV from the
   recombination dominated regime ($p/\pi^+ \sim 1$) to the fragmentation
   dominated regime ($p/\pi^+ \sim 0.1$).
   Right: the nuclear modification factor $R_{AA}$ for pions (solid),
   charged hadrons (dashed) and protons (dotted line). Data points are 
   $R_{AA}$ for $\pi^0$ from PHENIX \cite{Miod:02}.} 
\end{center}
\end{figure}

In Figure \ref{fig:3} we give the proton/pion ratio and the 
nuclear modification factor $R_{AA}$ for pions, protons and charged hadrons. 
The proton/pion ratio shows a plateau between 2 and 4 GeV/$c$ in accordance 
with experiment and a steep decrease beyond that. This decrease has not yet 
been seen and is a prediction of our work. 

In the quantity $R_{AA}$ the effect of jet quenching is manifest for all 
hadrons at large transverse momentum. For pions $R_{AA}$ grows only 
moderately at low $P_T$ in accordance with PHENIX data \cite{Miod:02}.
On the other hand recombination is much more important for protons, 
because they are suppressed in the fragmentation process. Therefore $R_{AA}$ 
nearly reaches a value of 1 below 4 GeV/$c$ for protons.

In summary, we have discussed recombination as a possible hadronization
mechanism in heavy ion collisions. We have shown that recombination can
dominate over fragmentation up to transverse momenta of 5 GeV/$c$.
Recombination provides a natural explanation for the baryon/meson
ratio and the nuclear suppression factors observed at RHIC.

{\bf Acknowledgments.}
This work was supported by RIKEN, Brookhaven National Laboratory, 
DOE grants DE-FG02-96ER40945 and DE-AC02-98CH10886, and by the 
Alexander von Humboldt Foundation.

\end{document}